\documentclass[12pt]{iopart}
\usepackage{times}

\usepackage{amssymb}
\usepackage{amsbsy}

\newcommand{\bsym}[1]{\boldsymbol{#1}}
\newcommand{\tb}[1]{\textbf{#1}}
\newcommand{\da}[1]{#1^{\,+}}

\begin{document}

\title{Quantization of the Linearized Kepler Problem}
\author{Julio Guerrero  and Jos\'e Miguel P\'erez }
\address{Depto. de Matem\'atica Aplicada, Universidad de Murcia, 30100 Murcia,
Spain}
\begin{abstract}

The linearized Kepler problem is considered, as obtained from the Kustaanheimo-Stiefel (K-S)
transformation, both for negative and positive energies. The symmetry group for the Kepler problem
turns out to be  $SU(2,2)$. For negative energies, the Hamiltonian
of Kepler problem can be realized as the sum of the energies of four harmonic oscillator with the
same frequency, with a certain constrain. For positive energies, it can be realized as the sum of the
energies of four repulsive oscillator with the same (imaginary) frequency, with the same constrain.
The quantization for the two cases, negative and positive
energies is considered, using group theoretical techniques and constrains. The case of zero energy is also
discussed.

\end{abstract}


\section{KS Regularization  of the Kepler problem.}

In this work we affront the task of the quantization of the Kepler problem,
given by the Hamiltonian defined on
$\mathbb{R}_0^3\times\mathbb{R}^3,\:\:$ 
${\cal H}=\frac{\vec{\mathbb{Y}}\cdot\vec{\mathbb{Y}}}
              {2m}-\frac{\gamma}{r},\: \hbox{where} \: r
              =\sqrt{\mathbb{X}^2},\:
              (\mathbb{X},\mathbb{Y})\in\mathbb{R}_0^3\times\mathbb{R}^3$.
For this purpose we shall use the linearization provided by
the KS regularization, as introduced by  P. Kustaanheimo and E. Stiefel, in
the spinorial version due to Jost (see \cite{kummer}).
The KS transformation regularizes the Kepler problem and linearizes it, showing
that the dynamical group of the Kepler problem is $SU(2,2)$.
The linearization means that the Kepler problem, for the case of negative energy, can be seen as a system of
\textbf{4 harmonic oscillators in resonance} subject to a constrain. For the case of positive energy, it turns to
be a system of  \textbf{4  repulsive harmonic oscillators in resonance} subject to
constrains. Finally, as we shall see, the singular case of zero energy can
be expressed as \textbf{4 free particles} subject to constrains.

The key point in the KS transformation is the commutativity of the diagram (see
\cite{kummer}):
\begin{equation} \label{diagramaconmutativo}
\begin{array}{ccc}
  (z,\,w)\,\in\,(I^{-1}(0))' & \stackrel{\mathfrak{C}}{\rightarrow}
  & (\eta,\,\zeta)\,\in\, I^{-1}(0)\subset \mathbb{C}^4 \\
  \pi\: \downarrow & \circlearrowleft & \downarrow\:\widehat{\pi} \\
  (\vec{x},\,\vec{y})\,\in\,\mathbb{R}^3_0\,\times\,\mathbb{R}^3
  & \stackrel{\nu^{-1}}{\rightarrow} &  (q,\,p)\,\in\,T^+S^3
\end{array}
\end{equation}

In this diagram $\nu$ is Moser transformation (see \cite{kummer}),
which allows us to see $T^+S^3$ as an embedded manifold in $\mathbb{R}^3_0\,\times\,\mathbb{R}^3$,
where $T^+S^3=\{(q,p)\,\in\,\mathbb{R}^{8},\,||q||=1,<p,q>=0,\,p\,\neq\,0\}$,
is named \textbf{Kepler manifold}.

The KS transformation is the map $\pi$, which can be seen as a symplectic lift of the Hopf fibration,
$\pi_0\,:\,\mathbb{C}^2_0  \rightarrow  \mathbb{R}^3_0, \quad
   z=(z_1,z_2)  \mapsto  \pi_0(z)\,:=\,<z,\bar{\sigma}z>$,
   ($\vec{\sigma}$ are Pauli matrices):
\begin{equation}
\hspace{-1cm} \pi:T^*\mathbb{C}^2_0  \rightarrow  T^*\mathbb{R}^3_0, \:
(z,w) \!  \mapsto \!  (\vec{x}=\pi_0(z),\vec{y}={\rm Im}<w,\vec{\sigma}z>/<z,z>),
\end{equation}
such that, $\pi^*\theta_{\mathbb{R}^3_0}=\theta_{\mathbb{C}^2_0}|_{(I^{-1}(0))'}=
2{\rm Im}<\!\!w,dz\!\!>$ (= $\theta_{\eta\,\zeta}=
{\rm Im}(<\!\!\eta,d\eta\!\!>-<\!\!\zeta,d\zeta\!\!>)$ up to a total differential) and $\theta_{\mathbb{R}^3_0}$ is the canonical potential
form restricted to $\mathbb{R}^3_0$.
The map $\mathfrak{C}=\frac{1}{\sqrt{2}}\left( \begin{array}{cc} \sigma_0 & \phantom{-}\sigma_0
\\ \sigma_0 & -\sigma_0 \end{array}\right)$ provides the injection of collision states.
The function $I =\frac{1}{2}(<\!\eta,\eta\!>-<\!\zeta,\zeta\!>)$ defines the regularized space $I^{-1}(0)$ 
($(I^{-1}(0))'$ doesn't contain collision states),
which is diffeomorphic to $\mathbb{C}^2_0\times S^3$
while $I^{-1}(0)/U(1)$ is diffeomorphic to $\mathbb{R}^3_0\times S^3$.

 The transformation, $\vec{\mathbb{X}}=\frac{1}{\sqrt{m}k}\vec{x},\quad
\vec{\mathbb{Y}}=k\sqrt{m}\vec{y}$, with $\rho=\sqrt{\vec{x}^2}$, relates the variables in the Kepler
problem to the variables used  in $\nu$.
The map $\widehat{\pi}$ is a symplectomorphism between $I^{-1}(0)/U(1)$
and $T^+S^3$, with the symplectic structures restricted to the corresponding
spaces.
The Kepler Hamiltonian for  negative energy
is associated with ${\cal J}=\frac{1}{2}(<\eta,\eta>+<\zeta,\zeta>)$, which corresponds to a system
of 4 harmonic oscillators in resonance 1-1-1-1.

We can proceed in the same manner for the case of positive energies changing the Kepler manifold by
$T^+H^n= \{(q,p)\::<q,q>=-1,\:<q,p>=0, \:p_0>||\vec{p}||\}$. For this
case,  the Kepler Hamiltonian is associated with $-\,P_0$ (see below), with the additional constrain
 $-\,P_0\,>\,0$ (besides $I^{-1}(0)$). However, due to its singular character
there is no possibility of considering the zero energy case in this
(geometrical) way. We shall use, in this case, a group theoretical
argument to study it.

The potential 1-form $\theta_{\eta\,\zeta}$ is left invariant by the Lie
group $U(2,2)$, which also leaves invariant the constrain $I$ when acting
on $\mathbb{C}^4_0$. This is thus the dynamical group for the Kepler
problem. A convenient basis for the Lie algebra $u(2,2)$ is given by the components of the momentum map associated with
its action on $\mathbb{C}^4_0$ (here $I$ is central):
\begin{equation}
\hspace{-2cm}
\begin{array}{l}
I,\quad {\cal J},\quad
\vec{M}=-\frac{1}{2}<\eta,\vec{\sigma}\,\eta>,\quad
\vec{N}=
\frac{1}{2}<\zeta,\vec{\sigma}\,\zeta>, \\
Q=(-Im<\eta,\zeta>,Re<\eta,\vec{\sigma}\zeta>),\quad
P=(Re<\eta,\zeta>,Im<\eta,\vec{\sigma}\zeta>).
\end{array}
\label{su22}
\end{equation}
\begin{center}
\begin{tabular}{|c|}
\hline
\textbf{
Table I: KS regularization with physical constants} \\ \hline
$\vec{\mathbb{X}}=\frac{1}{\sqrt{m}\,k}\,(\vec{Q}-\vec{R}'),\qquad
\vec{\mathbb{Y}}=k\,\sqrt{m}\,\frac{\vec{P}}{||P||+P_0},$
\\ \hline   \vspace{-0.4cm}   \\
 ${\cal H}=\frac{\vec{\mathbb{Y}}^2}{2m}-\frac{\gamma}{||\vec{\mathbb{X}}||}
 =\frac{k}{2(||P||+P_0)}(k(||P||-P_0)-2\gamma\sqrt{m}),$ 
 \quad
   $\vec{\cal AM}=\vec{L}=\vec{\mathbb{X}}\times\vec{\mathbb{Y}}=\vec{M}+\vec{N},$ 
   \\
   $\vec{R'}=\vec{M}-\vec{N},$\qquad  
   $\vec{{\cal RL}}=
   \frac{\vec{\mathbb{Y}}\times\vec{L}}{m}-\gamma\frac{\mathbb{X}}{||\mathbb{X}||}=
   \frac{\vec{R'}(kP_0+\gamma\sqrt{m})+\vec{Q}(k||P||-\gamma\sqrt{m})}
   {\sqrt{m}(||P||+P_0)}.$\\
 \hline
\end{tabular}
\end{center}

\section{Quantization of  Kepler problem: $E\,<\,0$.}

The KS transformation reveals that the Kepler problem for negative
energies can be seen as the Hamiltonian system
$(\mathbb{C}^4,\theta_{(\eta,\zeta)},{\cal J})$ restricted to $I^{-1}(0)$.
Defining $\tb{C}=(\tb{C}_1,\tb{C}_2)=(\eta,\da{\zeta}),\,\,
\tb{C}_i\in\mathbb{C}^2$, the Hamiltonian ${\cal J}$ adopts the form
$\mathcal{H}_{\rm har}=\omega\tb{C}\da{\tb{C}}$ which corresponds to
four harmonic oscillators. The quantization of this system can be
obtained from the group law of the corresponding symmetry group (a central extension of it by $U(1)$,
rather, see \cite{aa82}):
\begin{equation}
\begin{array}{l}
\lambda''=\lambda'+\lambda, \qquad
\tb{C}''=\tb{C}'\,e^{-\,i\,\lambda}+\tb{C}, \quad
\da{\tb{C}''}=\da{\tb{C}'}\,e^{i\,\lambda}+\da{\tb{C}},\\
\varsigma''=\varsigma'\,\varsigma\,\exp[\frac{i}{2}\,(i\,\tb{C}'\,\da{\tb{C}}\,e^{-\,i\,\lambda}
-i\,\da{\tb{C}'}\,\tb{C}\,e^{i\,\lambda})],
\end{array}
\label{Oscilador}
\end{equation}\noindent
where $\tb{C},\da{\tb{C}}\in \mathbb{C}^4$, $\varsigma\in U(1)$ and $\lambda=\omega\,t\:\in\:\mathbb{R}$.
We can obtain the quantum version of this system using any
geometrical (like Geometric Quantization, see \cite{GQ})  or group-theoretical
method, like Group Approach to Quantization (GAQ, see \cite{aa82}), the one used here.

The resulting wave functions (defined on the group) are
$\psi=\varsigma e^{-\frac{1}{2}\bsym{C}\da{\bsym{C}}}\phi(\da{\tb{C}},\lambda)$,
and Schr\"odinger equation for this system is $i\frac{\partial \phi}{\partial
\lambda}=i\da{\tb{C}}\frac{\partial \phi}{\partial \da{\bsym{C}}}$.
In this formalism, quantum operators are constructed from the
right-invariant vector fields on the group (\ref{Oscilador}), and in this
case \textbf{creation} and \textbf{annihilation} operators are given
by $\da{\widehat{\tb{C}}}=X^R_{\bf C}$ and
$\widehat{\tb{C}}=X^R_{\da{\bf C}}$, respectively.
Since the momentum map (\ref{su22}) is expressed as quadratic functions
on $\tb{C}$ and $\da{\tb{C}}$, we can resort to Weyl prescription to
obtain the quantization of these functions on the (right) enveloping algebra of
the group (\ref{Oscilador}). In this way we obtain a Lie algebra of quantum operators
isomorphic to the one satisfied by the momentum map (\ref{su22}) with the
Poisson bracket associated with $\theta_{\eta\,\zeta}$. The Hamiltonian
operator and the quantum version of the constrain, when acting on wave functions
are given by (${\cal W}=\varsigma e^{-\frac{1}{2}{\bf C}\cdot\da{{\bf C}}}$):
$\widehat{J}\psi=-\frac{1}{2}{\cal W}(2+
   \da{\tb{C}}\frac{\partial}{\partial \da{\bsym{C}}})\phi$  and
   $\widehat{I}\psi=-\frac{1}{2}{\cal W}(
      \da{\tb{C}}_1\frac{\partial}{\partial \da{\bsym{C}}_1} -
      \da{\tb{C}}_2\frac{\partial}{\partial \da{\bsym{C}}_2})\phi$\,.
To obtain the quantum version of the Kepler manifold (that is, the Hilbert
space of states of the Hydrogen atom for $E<0$), we must
impose the constrain $\widehat{I}\psi=0$. This means that the
energy of the first two oscillators must equal the energy of the other
two. It is easy to check that the operators in the (right) enveloping algebra of
the group (\ref{Oscilador}) preserving the constrain (see \cite{frachall,modular}
for a characterization of these operators) is the
algebra $su(2,2)$ of the quantum version of the momentum map
(\ref{su22}). These operators act irreducibly on the constrained Hilbert
space, as can be checked computing the Casimirs of $su(2,2)$, which are
constant.
The quantum operators commuting with the Hamiltonian (and providing
the degeneracy of the spectrum) are $\widehat{\vec{M}}$ and $\widehat{\vec{N}}$.
They define two commuting $su(2)$ algebras in the same representation
($(\widehat{\vec{M}})^2=(\widehat{\vec{N}})^2=\frac{1}{4}\,(\widehat{{\cal
J}})^2-\frac{1}{4}$), and linear combinations of them provide us
with the angular momentum and the Runge-Lenz vector (see Table I).

The relation between the Kepler Hamiltonian ${\cal H}$ and the
Hamiltonian ${\cal J}$ is ${\cal H}=-\,\frac{m\,\gamma^2}{{2\,\cal
J}^2}$. If we act on eigenstates of the number operator for each
oscillator,
$\psi_{n_1,n_2,n_3,n_4}\,\approx\,
(\da{C}_{11})^{n_1}(\da{C}_{12})^{n_2}(\da{C}_{13})^{n_3}(\da{C}_{14})^{n_4}$, and
taking into account that:
$\widehat{\mathcal{E}}\psi_{n_1,n_2,n_3,n_4}=
\widehat{{\cal J}}\psi_{n_1,n_2,n_3,n_4}=\frac{1}{2}(2+\sum n_i)\psi_{n_1,n_2,n_3,n_4}$,
we recover the \textbf{spectrum of the Hydrogen atom}, $E_{n}=-\,\frac{m\,\gamma^2}{2\,n^2},\:\:
n=1+n_1+n_2$. The degeneracy is provided by the dimension of the representations of the algebra
$su(2)\times su(2)$, which turn to be $n^2$ (if spin 1/2 is considered, the degeneracy is doubled).

\section{Quantization of  Kepler problem: $E\,>\,0$.}

The KS transformation, for the case of positive energies, maps the Kepler Hamiltonian to
the function $-\,P_0$, with the constrains $I=0$ and $-\,P_0\,>\,0$
and with the same potential 1-form $\theta_{(\eta,\zeta)}$.
Performing the change of variables:
\begin{displaymath}
\hspace{-2.5cm} q_i=\frac{1}{2}(\alpha_{i+1}+\nu_{i+1}),\,
i=\,0,1,2,3,\,\,
 p_i=\frac{1}{2}(\alpha_{i+2}-\nu_{i+2}),\,
i=0,2,\,\, p_i=\frac{1}{2}(\nu_{i}-\alpha_{i}),\,
i=1,3
\end{displaymath}
$z_1=q_0+i\,q_1$, $z_2=q_2+i\,q_3$,
$w_1=p_0+i\,p_1$ and $w_1=p_2+i\,p_3$, the new Hamiltonian
$-\,P_0$ can be written as $\sim\,\boldsymbol{ \alpha\,\nu}$, which corresponds to four
repulsive oscillators in resonance 1-1-1-1. The (extended) symmetry group
for this system is given by (see \cite{aaw}):
\begin{equation}
\begin{array}{l}
\lambda''=\lambda'+\lambda, \qquad
\boldsymbol{\alpha}''=\boldsymbol{\alpha}'\,e^{\lambda}+\boldsymbol{\alpha}, \quad
\bsym{\nu}''=\bsym{\nu}'\,e^{-\,\lambda}+\bsym{\nu},\\
\varsigma''=\varsigma'\,\varsigma\,\exp[\frac{i}{2}\,(\bsym{\alpha\,\nu}'\,e^{-\,\lambda}
-\bsym{\alpha}'\,\bsym{\nu}\,e^{\lambda})],
\end{array}
\label{repulsivo}
\end{equation}     \noindent
where $\bsym{\alpha,\nu}\in \mathbb{R}^4$ and $\lambda\,:=\,\omega\,t\:\in\:\mathbb{R}$.
The Hamiltonian for this system is $\mathcal{H}_{rep}=-\,\omega\,\bsym{\alpha\,\nu}$.
Applying GAQ we obtain that the wave functions are
$\psi=\varsigma\,e^{\frac{-\,i\,\bsym{\alpha\,\nu}}{2}}\,\phi(\bsym{\nu},\lambda)$,
and the Schr\"odinger
equation is $i\,\frac{\partial \phi}{\partial
\lambda}\,=i\,\bsym{\nu}\,\frac{\partial \phi}{\partial \bsym{\nu}}$.

With the same procedure as in the case of negative energies, we construct a realization of the algebra
(\ref{su22}) resorting to the enveloping algebra of the group (\ref{repulsivo}). The
quantum version of $P_0$ and the constrain are (${\cal W}=\varsigma e^{-\frac{i}{2}\,\bsym{\alpha\nu}}$):
The representation defined by the quantum version of (\ref{su22}) is
irreducible, since the Casimirs are constant. 
The additional constrain $-\widehat{P_0}>0$ restrict
further the algebra of physical operators, being generated by the
\textbf{angular momentum} $\widehat{\vec{L}}$ and the \textbf{Runge-Lenz} vector $\widehat{\vec{Q}}$.
These operators close the \textbf{Lorentz algebra},
$[\widehat{L}_i,\,
\widehat{Q}_j]=-\,2\,i\,\epsilon_{ijk}\,\widehat{Q}_k$,
$[\widehat{L}_i,\,
\widehat{L}_j]=-\,2\,i\,\epsilon_{ijk}\,\widehat{L}_k$ y
$[\widehat{Q}_i,\,
\widehat{Q}_j]=+\,2\,i\,\epsilon_{ijk}\,\widehat{L}_k$, as expected.

Again, using the relation between the Kepler Hamiltonian and $P_0$,
${\cal H}=\frac{m\,\gamma^2}{2\,P_0^2}$, we obtain the spectrum of the
Hydrogen atom for positive energies,
$E_{\tau}=\frac{m\,\gamma^2}{2\,\tau^2}\:;\:\:\tau=2+\tau_1+\tau_2$.

\section{Quantization of  Kepler problem: $E=0$.}

For the case of zero energy, it is not clear which is the Kepler manifold, and some of the expressions
obtained for negative and positive energies have no limit when the energy
(related to $k$) goes to zero. We propose a candidate for the linearized zero
energy Kepler problem, using a group-theoretical argument. The idea is to
look, in the $su(2,2)$ algebra, for an appropriate Runge-Lenz vector $\vec{{\cal RL}}$ 
(angular momentum doesn't change, but Runge-Lenz vector depends on the energy).
From Table I, we observe that
${\cal H}=-\frac{k^2}{2}\mapsto
\vec{{\cal RL}}=\frac{k}{\sqrt{m}}\vec{R}'$ and  ${\cal H}=\frac{k^2}{2}\mapsto
\vec{{\cal RL}}=\frac{k}{\sqrt{m}}\vec{Q}$. This suggests the choice, for zero energy
(and for $k\,\neq\,0$),
$\vec{\cal RL}=\frac{k}{2\,\sqrt{m}}\,(\vec{R}'+\vec{Q})$,
implying that the Hamiltonian is the sum of the two Hamiltonian (with equal opposite energies!),
$\mathbf{E}_0=\frac{1}{2}\,({\cal J}-P_0)=\frac{1}{2}\,(||P||-P_0)$.  This
can be achieved if we impose the constrain (zero energy)
$||P||-P_0=\frac{2\,\gamma\,\sqrt{m}}{k}$, derived from our choice of Runge-Lenz vector. 
The constrain $I=0$
is also satisfied in this case, by construction.
Now we perform the following change of variables:
\begin{displaymath}
\hspace{-2cm}
b_0=q_0, \quad a_0=q_1,\quad B_0=p_1, \quad A_0=p_0, \quad
b_1=q_2, \quad a_1=q_3, \quad B_1=p_3,  \quad A_1=p_2,
\end{displaymath}
obtaining that the Hamiltonian is written
$\mathbf{E}_0=\frac{1}{2}\sum_i(A_i^2+B_i^2)$, that is, a system of \textbf{four
free particles}. It must be stressed that the variables $a_i$, $A_i,\,i=0,1$
satisfy commutation relations with opposite sign to that of $b_i$, $B_i,\,i=0,1$,
as can be seen from the potential 1-form:
\begin{equation}
\hspace{-1.5cm}
\theta=-B_0db_0+b_0\,dB_0-B_1db_1+b_1dB_1
+A_0da_0-a_0dA_0+A_1da_1-a_1dA_1\,.
\end{equation} \noindent
The (extended) symmetry group for the system of four free particles is given by:
\begin{displaymath}
\hspace{-2.35cm}
\begin{array}{l}
\lambda''=\lambda'+\lambda,\quad
\tb{a}''=\tb{a}+\tb{a}'+\tb{A}'\lambda,\quad
\tb{b}''=\tb{b}+\tb{b}'+\tb{B}'\lambda,\quad
\tb{A}''=\tb{A}+\tb{A}',\quad
\tb{B}''=\tb{B}+\tb{B}'\\
\varsigma''=\varsigma'\,\varsigma\,\exp\{i\,[(\tb{a}'\,\tb{A}+\lambda\,(\tb{A}'\,\tb{A}
+\frac{1}{2}\,\tb{A}^{'\,2})]\}\,\exp\{i\,[(\tb{b}\,\tb{B}'+\frac{1}{2}\,\tb{B}^{'\,2}\,\lambda]\},
\end{array}
\end{displaymath}
where the relative sign in the canonical structure of the $(a_i,A_i)$ and
$(b_i,B_i)$ variables has been taken into account in the
2-cocycle.
Repeating the procedure of the non-zero energy cases, we obtain
a new realization of the algebra $su(2,2)$ on this space, which is irreducible. 

The energy and constrain operators are
($\mathcal{W}= \varsigma e^{-i{\bf b}\cdot {\bf B}}e^{\frac{i}{2}
({\bf B}^2-{\bf A}^2)\lambda}$):
$\widehat{\mathbf{E}}_0\,\psi=
-\frac{1}{2}\mathcal{W}\sum_i(A_i^2+B_i^2)\varphi$, and
$\widehat{I}\,\psi=i\mathcal{W}\,\sum_i(A_i\frac{\partial}{\partial B_i}-
B_i\frac{\partial}{\partial A_i})\varphi.$
If we impose $E=0$, we obtain the new constrain
$\sum_i(A_i^2+B_i^2)  =\frac{2\gamma\sqrt{m}}{k}$.
Therefore $k$ defines a foliation in spheres. The operators preserving this new constrain are
again the angular momentum $\widehat{\vec{L}}$ and the Runge-Lenz vector $\widehat{\vec{S}}=
\widehat{\vec{R}'}+\widehat{\vec{Q}}$,
closing \textbf{the Euclidean algebra $e(3)$}, as expected.
This guarantees that we have made the correct choice of Hamiltonian and Runge-Lenz vector for $E=0$.

 An interesting application of these results is the fact that the
linearization is preserved in some perturbed problems, such as the lunar
problem or the Stark effect, see \cite{enpreparacion}.

A similar study can be found in \cite{Gracia-Bondia}, where the quantization of the Kepler problem
for $E\neq 0$ is considered in the Weyl-Wigner-Moyal formalism using the KS
transformation.


\section*{References}

\begin{thebibliography}{99}
        \bibitem{kummer} M. Kummer,
        Comm. Math. Phys. \textbf{84},  133 (1982).

      \bibitem{GQ} N.M.J. WoodHouse,
        \textit{Geomeric Quantization (2$^{nd}$ ed.)},
        Oxford University Press (1991).

        \bibitem{aa82} V. Aldaya and J.A. de Azc\'arraga,
        J. Math. Phys. \textbf{23}, 1297 (1982).

        \bibitem{aaw} V. Aldaya, A. de Azc\'arraga and K.B.Wolf,
        J. Math. Phys. \textbf{25}, 506 (1984).

         \bibitem{frachall} V. Aldaya, M. Calixto, J. Guerrero,
        Comm. Math. Phys. \textbf{178}, 399 (1996).

         \bibitem{modular} J. Guerrero, M. Calixto and V. Aldaya,
        J. Math. Phys. \textbf{40} 3773 (1999).

        \bibitem{enpreparacion} J. Guerrero and J.M. P\'erez, in
        preparation.

        \bibitem{Gracia-Bondia} J.M. Gracia-Bondía, Phys. Rev. {\bf A30}, 691 (1984)

\end{thebibliography}
\end{document}